\documentclass[aps,pre,preprint,superscriptaddress]{revtex4-1}

\usepackage{amsmath} 
\usepackage{amssymb} 
\usepackage{amsfonts}

\usepackage{color}
\usepackage[]{graphicx}

\newcommand{\kB}{k_B}
\newcommand{\tens}[1]{#1}

\renewcommand{\vec}[1]{\mathbf{#1}}
\newcommand{\mrm}[1]{\mathrm{#1}}

\newlength{\figurewidth}
\begin{document}

\title{Nonequilibrium Structure of Colloidal Dumbbells under Oscillatory Shear}

\author{Nils Heptner}
\affiliation{{Institut f\"{u}r Weiche Materie und funktionale Materialien, Helmholtz-Zentrum Berlin f\"{u}r Materialien und Energie GmbH, Hahn-Meitner Platz 1, 14109 Berlin, Germany}}
\affiliation{{Institut f{\"u}r Physik, Humboldt-Universit{\"a}t zu Berlin, Newtonstr.~15, D-12489 Berlin, Germany}}
\author{Fangfang Chu}
\affiliation{{Institut f\"{u}r Weiche Materie und funktionale Materialien, Helmholtz-Zentrum Berlin f\"{u}r Materialien und Energie GmbH, Hahn-Meitner Platz 1, 14109 Berlin, Germany}}
\affiliation{{Institut f{\"u}r Physik, Humboldt-Universit{\"a}t zu Berlin, Newtonstr.~15, D-12489 Berlin, Germany}}
\author{Yan Lu}
\affiliation{{Institut f\"{u}r Weiche Materie und funktionale Materialien, Helmholtz-Zentrum Berlin f\"{u}r Materialien und Energie GmbH, Hahn-Meitner Platz 1, 14109 Berlin, Germany}}
\author{Peter Lindner}
\affiliation{Institut Laue-Langevin, 71 avenue des Martyrs, 38042 Grenoble Cedex 9, France}
\author{Matthias Ballauff}
\affiliation{{Institut f\"{u}r Weiche Materie und funktionale Materialien, Helmholtz-Zentrum Berlin f\"{u}r Materialien und Energie GmbH, Hahn-Meitner Platz 1, 14109 Berlin, Germany}}
\affiliation{{Institut f{\"u}r Physik, Humboldt-Universit{\"a}t zu Berlin, Newtonstr.~15, D-12489 Berlin, Germany}}
\author{Joachim Dzubiella}
\email{joachim.dzubiella@helmholtz-berlin.de}
\affiliation{{Institut f\"{u}r Weiche Materie und funktionale Materialien, Helmholtz-Zentrum Berlin f\"{u}r Materialien und Energie GmbH, Hahn-Meitner Platz 1, 14109 Berlin, Germany}}
\affiliation{{Institut f{\"u}r Physik, Humboldt-Universit{\"a}t zu Berlin, Newtonstr.~15, D-12489 Berlin, Germany}}

\begin{abstract}

We investigate the nonequilibrium behavior of dense, plastic-crystalline suspensions of mildly anisotropic colloidal hard dumbbells under the action of an oscillatory shear field by employing  Brownian dynamics computer simulations. In particular, we extend previous investigations, where we uncovered novel nonequilibrium phase transitions,  to other aspect ratios and to a larger nonequilibrium parameter space, that is, a wider range of strains and shear frequencies. We compare and discuss selected results in the context of novel scattering and rheological experiments.  Both simulations and experiments demonstrate that the previously found transitions from the plastic crystal phase with increasing shear strain also occur at other aspect ratios. We explore the  transition behavior in the strain-frequency phase  and summarize it in a nonequilibrium phase diagram. Additionally, the experimental rheology results hint at a slowing down of the colloidal dynamics with higher aspect ratio.

\end{abstract}

\maketitle

\section{Introduction}

The equilibrium and nonequilibrium behavior of colloidal hard-sphere suspensions has been investigated extensively~\cite{Pusey1989,Ackerson1991,Ackerson1990,Ackerson1983} as it constitutes the simplest model system to gain a better understanding of the structure and phase behavior of colloidal suspensions in some of their fundamental aspects.
Notwithstanding that the particles interact isotropically by excluded volume only, hard sphere colloids show a nontrivial phase diagram already in equilibrium and glassy behavior above a certain volume fraction~\cite{Pusey1986,Pusey1989,Pusey2009}.
Regarding their nonequilibrium behavior, hard sphere suspensions subjected to shear have been of particular interest, since shear is one of the most common external fields~\cite{Besseling2012,Ackerson1988}.
Here, many interesting forms of order-disorder transitions have been observed in experiments and simulations under various shear conditions~\cite{Besseling2012,Ackerson1988}. 
These include steady shear (constant rate)~\cite{Ackerson1988}, oscillatory shear (fixed strain amplitude and frequency)~\cite{Ackerson1988,Besseling2012}, and oscillatory fixed shear rate protocols~\cite{Ackerson1990a}. 

In particular, various nonequilibrium states have been identified in spherical colloids under shear and have been well characterized up to now using a combination of scattering methods~\cite{Ackerson1988}, optical techniques~\cite{Haw1998,Besseling2012}, simulations~\cite{Besseling2012} and theoretical models~\cite{Loose1994}. 
Similar order-disorder transitions of colloids under shear have been reported for clusters of soft spheres~\cite{Nikoubashman2012} and highly charged spheres~\cite{Yan1994,Biehl2004,Palberg2002}.
Most recently, \citeauthor{Besseling2012} combined confocal microscopy and Brownian dynamics ({BD}) simulations to explore the nonequilibrium behavior of colloidal hard spheres under oscillatory shear~\cite{Besseling2012}.
Here, for small strains a face-centered cubic ({FCC}) twin is found to be stable corroborating with the classical results~\cite{Ackerson1988}.
At high strains the predominant structure is found to be registered sliding of hexagonal close packed ({HCP}) layers.
Moreover, a dense direction of the {HCP} layer prefers to be parallel to the velocity at high strain amplitudes, whereas at low strains a dense direction is parallel to the flow direction.
This behavior results in a $30^{\circ}$ turn in the scattering pattern~\cite{Haw1998}.
The corresponding diffraction patterns~\cite{Ackerson1990a,Ackerson1988} show three-fold symmetries.
\citeauthor{Besseling2012} have calculated an extensive nonequilibrium state diagram for hard spheres under oscillatory strain, which categorizes further high-strain structures.

\begin{figure}[ht]
	\centering
	\includegraphics[width=\figurewidth]{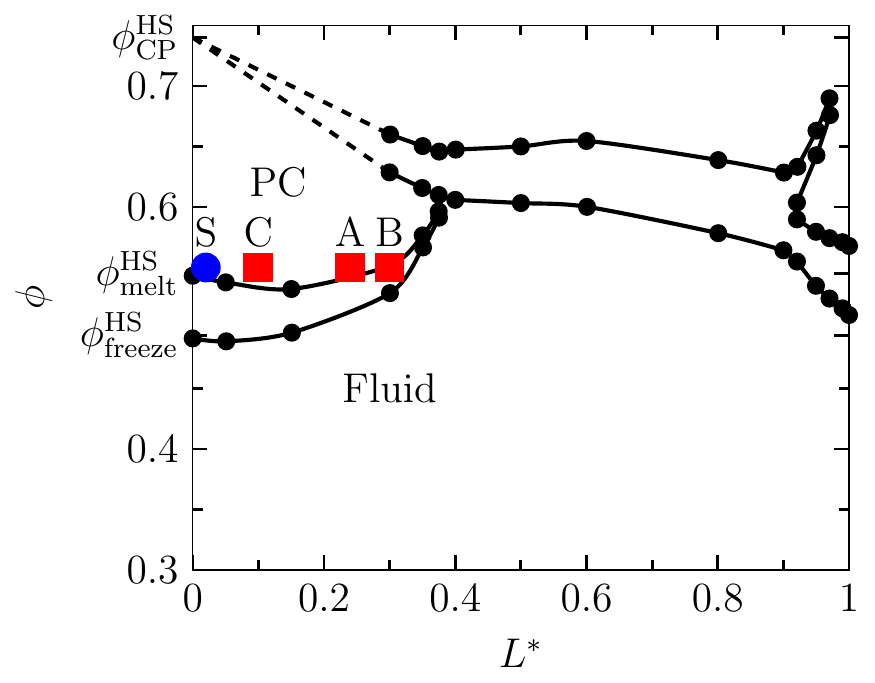}
	\caption{(Color online) Hard dumbbell phase diagram in  the volume fraction $\phi$ to aspect ratio $L^\ast$ plane~\cite{Marechal2008}. 
	The state points for dumbbells A ($\phi= 0.55$, $L^\ast=0.08$), B  ($\phi= 0.55$, $L^\ast=0.24$), and C ($\phi= 0.55$, $L^\ast=0.3$), considered in this work are marked by symbols (\textcolor{red}{$\blacksquare$}). The state point~S (\textcolor{blue}{$\bullet$}) denotes an almost hard-sphere-like reference system ($\phi= 0.55$, $L^\ast=0.02$). For better orientation, the freezing ($\phi^{\rm HS}_{\rm freeze}$), melting ($\phi^{\rm HS}_{\rm melt}$) and close packed ($\phi^{\rm HS}_{\rm CP}$) packing densities of the {HS} system are indicated at the vertical axis.}
	\label{fig:phasediagram}
\end{figure}

In general, however, colloids are anisotropic and as such there is a rising interest in the behavior of suspensions of non-spherical particles for fundamental understanding~\cite{Blaaderen2006,Hansen-Goos2010,Eisenriegler2005} or applications, such as constituents for novel materials~\cite{Glotzer2007} or photonics~\cite{Demirors2010,Liu2014,Hosein2010}.
Desirable on a fundamental level are hard particles with slight anisotropy that weakly perturb the isotropic interactions of spherical reference systems.
One popular experimental realization is a system of steeply repulsive \emph{'dumbbells'}, that is a colloidal dimer made up by two fused equally-sized hard spheres~\cite{Mock2007,Mock2007a}.
The equilibrium phase diagram and the stability and nucleation processes of (plastic) crystal phases of hard dumbbells have been mapped out comprehensively by means of Monte Carlo ({MC}) simulations~\cite{Vega1992a,Vega1997,Marechal2008,Ni2011}.
Plastic crystal phases are characterized by a crystalline center of mass order and a lack of long-range order of the particles' orientations~\cite{Wojciechowski1991,Frenkel1984,Lynden-Bell1994}, in which the latter is the distinguishing attribute in comparison to a fully ordered crystal.
For the convenience of the reader the relevant parts of the phase diagram are replotted in Fig.\nobreakspace \ref {fig:phasediagram}.

Out of equilibrium, the impact of the weak anisotropy and its accompanying translational-rotational coupling~\cite{Lynden-Bell1994} on the rheology and the structure of dispersions under shear has been of keen interest.
This aspect has been investigated by means of mode-coupling theory~\cite{Zhang2009,Zhang2012}, neutron scattering~\cite{Mock2007} and rheological experiments~\cite{Kramb2010,Kramb2011a}.
A few years ago, we have presented a very neat experimental realization~\cite{Chu2012} of monodisperse hard dumbbells (aspect ratios $0.24$ and $0.30$), which matches the plastic crystal phase boundaries well in the estimated phase diagram (Fig.\nobreakspace \ref {fig:phasediagram}). 
Employing this well-defined experimental model system, we have, in fact, very recently shown by a combination of rheology-scattering (rheo-SANS) and Brownian dynamics (BD) simulations that hard plastic-crystalline dumbbells undergo nonequilibrium transitions under oscillatory shear similar than hard spheres but the nature of the transition differs strikingly~\cite{Chu2015}.
In particular, we observed a finite orientational correlation arising on increasing strain amplitudes~\cite{Chu2015} and a more vigorous transition with large implications on rheology and the yielding behavior. 

In this work, we follow up on our previous work and extend our computational work to other aspect ratios and a wider frequency range.
Our focus is thus the region of small to moderate elongations (aspect ratios $L^\ast < 0.4$) and high volume fractions where the plastic crystal ({PC}) phase predominates~\cite{Vega1992}.
Furthermore, kinetics properties of the suspensions are investigated by our simulations. 
We also present new rheology experiments on a second aspect ratio ($0.3$). 
We note that the aspect ratio of $0.3$ represents an anisotropy very close to that of a nitrogen molecule, that also features a plastic crystal phase~\cite{Mills1975,Mills1986}, referred to as $\beta$-phase, so one could consider our system as \emph{colloidal nitrogen}.

\section{Methods}

\subsection{Brownian Dynamics}

The {BD} simulations are carried out using Ermak's~\cite{Ermak1975} method for interacting particles in solution with an additional term to account for the oscillatory shear force.
We neglect hydrodynamic interactions among the particles as the systems under consideration are disturbed in a regime where the driving forces are in the order of the viscous forces, i.e., the P\'{e}clet numbers are small, see also related work~\cite{Besseling2012}.
The dumbbell particles are represented by a two-segment Yukawa model, in which the particles interact via two spherical beads constraint at the constant distance $L$ with a steep Yukawa potential (decay length $\kappa^{-1} = 0.05\sigma$) defined by 
\begin{align}
	V(r) &= \epsilon \frac{\sigma}{r} \exp\left\{-\kappa\left(r - \sigma\right)\right\},
	\label{eq:yukawa}
\end{align}
where $\epsilon = k_B T$ and $\sigma$ set the energy and length scales, respectively, and $r$ denotes the center-to-center distance between two beads.
The parameter $\kappa$ tunes the softness of the interaction and is chosen to maintain computational performance and resemble hard particle behavior.
We employ a forward Euler scheme in order to integrate the equations of motion~\cite{Ermak1975,Loewen1994}. 
The time scale is set to the Brownian time $\tau = \sigma^2 / D_0^S$ of a single bead of diameter $\sigma$ and diffusivity $D_0^S = k_B T(3\pi\eta_s\sigma)^{-1}$ with a solvent viscosity $\eta_s$.
The parallel and perpendicular center of mass ({COM}) coordinates are updated according to 
\begin{align}
	\label{eq:update_par}
	\vec{R}_{i,\parallel}^{n+1} &= \vec{R}_{i,\parallel}^{n} + \Delta t \frac{D_{\parallel} }{\kB T} \vec{F}^n_{i,\parallel} + \delta r_{i,\parallel} \vec{u}^n_{i},\\
	\label{eq:update_perp}
	\vec{R}_{i,\perp}^{n+1} &= \vec{R}_{i,\perp}^{n} + \Delta t \frac{D_{\perp} }{\kB T} \vec{F}^n_{i,\perp} + \delta r_{i,1}\vec{e}^n_{i,1} + \delta r_{i,2} \vec{e}^n_{i,2},\\
	\label{eq:update_com}
	\vec{R}_i^{n+1} &= \vec{R}_{i,\parallel}^{n+1} + \vec{R}_{i,\perp}^{n+1} + \Delta t \dot\gamma(t) R_y^n \vec{e}_x.
\end{align}
The shear flow only affects the {COM} transport in $x$-direction through the last term in equation (Eq.\nobreakspace \textup {(\ref {eq:update_com})}).
The directors are updated following 
\begin{align}
	\vec{u}_i^{n+1} &= \vec{u}_i^{n} + \Delta t \frac{D_r}{\kB T} \vec{T}_i^n \times \vec{u}^n_i + \delta x_1 \vec{e}_{i,1}^n  + \delta x_2 \vec{e}_{i,2}^n,
\end{align}
where $\vec{T}_i^n$ is the total torque exerted on particle $i$ at time $t = n\Delta t$. 
The single-particle diffusion coefficients parallel $D_\parallel$, perpendicular $D_\perp$ to the long axis and $D_r$ for the rotation about the short axis depend on the particle geometry, their respective values are listed below (Table\nobreakspace \ref {tab:diffusion}).
The torque is comprised of the inter-particle and background-flow contributions, via
\begin{align}
	\vec{T}_i(t) &= \vec{T}_i^p(t) - \frac{k_B T}{D_r}\left\lbrace \vec{u}_i(t)\times\tens{\Gamma}(t)\cdot\vec{u}_i(t)\right\rbrace.
\end{align}

	\begin{figure}[ht]
		\centering
		\includegraphics[width=\figurewidth]{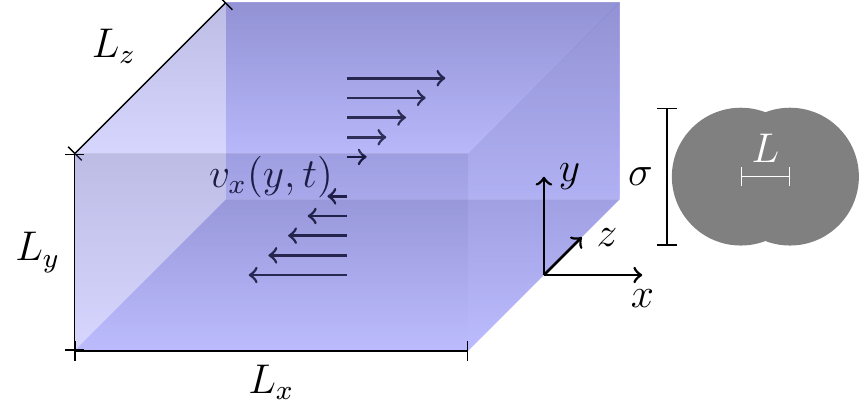}
		\caption{(Color online) Sketch of the imposed shear flow and the used coordinate system where the velocity is in $x$, the velocity gradient in $y$, and the vorticity is along the $z$-direction. The dumbbells' beads diameter is $\sigma$ and the aspect ratio (elongation) is defined as $L^\ast = L/\sigma$, where $L$ is the center-to-center distance of both beads. In the simulations, we apply Lees-Edwards boundary conditions~\cite{Lees1972}.}
		\label{fig:sketch_system}
	\end{figure}

We impose a time-dependent linear shear flow in the $x$-direction, such that the shear gradient is parallel to $\vec{e}_y$ and the vorticity is in $z$-direction.
As the flow velocity $\vec{v}(y) = v_x(y) \vec{e}_x$ vanishes at $y = 0$ and depends linearly on $y$, the velocity-vorticity plane is the plane of (spatially) constant flow velocity.
Hence, the sheared system retains symmetry in this particular plane, cf.~Fig.\nobreakspace \ref {fig:sketch_system}. 
Mathematically, the shear flow is thus described by the rate-of-strain tensor
\begin{align}
	\tens{E}(t) &= \dot\gamma(t)
							\begin{pmatrix}
								0 & 1 & 0 \\
								0 & 0 & 0 \\
								0 & 0 & 0 
							\end{pmatrix},
\end{align}
which may be written as a sum of symmetric and anti-symmetric tensors, i.e., the shear and vorticity tensors
\begin{align}
	\tens{E}(t) = \tens{\Gamma}(t) &+ \tens{\Omega}(t) \\
	= \frac{1}{2}\dot\gamma(t)%
								\begin{pmatrix}
									0 & 1 & 0 \\ 
									1 & 0 & 0 \\
									0 & 0 & 0 
								\end{pmatrix}
	&+ \frac{1}{2}\dot\gamma(t)%
								\begin{pmatrix}
									0 & 1 & 0 \\ 
									-1 & 0 & 0 \\
									0 & 0 & 0 
								\end{pmatrix}.
\end{align}
The symmetric part $\tens{\Gamma}$ describes the pure elongational flow, while the anti-symmetric part $\tens{\Omega}$ describes the vorticity of the flow. 
The vorticity is spatially uniform and quantifies the solvent contribution to the rotation of the particles.
The time-dependent strain with the dimensionless amplitude $\gamma_\mrm{max}$ imposed by the flow field on the suspension is thus given by
\begin{align}
	\gamma(t) &= \gamma_{\mathrm{max}} \sin(\omega t).
	\label{eq:oscillatory_strain}
\end{align}
Hence, the linear solvent velocity profile is
\begin{align}
	v_x(y, t) &= \dot\gamma(t) y .
	\label{eq:velocity_profile}
\end{align}
In order to compare the driving force to the intrinsic viscous forces, we define the P\'{e}clet numbers
\begin{align}
	&\mrm{Pe} = \frac{1}{12}f\gamma_\mathrm{max} \sigma^2 / D_0^S = \frac{1}{12} f \gamma_\mathrm{max} \tau,\ \text{and} &\\
	&\mrm{Pe}_r = 2\pi f \gamma_\mathrm{max} / D_r,&
\end{align}
where the latter is a definition respecting the time scale set by the rotational Brownian motion.
The maximum shear rate $\dot\gamma_\mrm{max}=\gamma_\mrm{max}/\tau$ sets the time scale of the driving force exerted by the shear flow.

In order to compare the results of our simulations, the time scales set by the diffusion constants and sizes of the particles are important.
For dumbbells with $L^\ast \approx 0.24$ the parallel ($D_\parallel = 0.93 D_0^S$), perpendicular ($D_\perp = 0.89 D_0^S$) and rotational ($D_r = 0.69 D_r^S = 0.69 (3 D_0^S) / (2 R_H)^2$) diffusivities have been obtained by the shell bead model method (SHM) and matched to the experimental data described in our previous work~\cite{Chu2012}.
In Table\nobreakspace \ref {tab:diffusion} the reader may find the exact parameters used for the respective systems.
Thus, the rotational Brownian time scale is $\tau_r = 0.69 (2R_H)^2 / (3D_0^S)$ in this case.
In length and time units of $2 R_H$ and $\tau$ the translational P\'{e}clet number is defined as $\mrm{Pe} = \frac{1}{12} f \gamma_\mathrm{max} (2 R_H)^2 / D_0^S$, where $R_H$ is the hydrodynamic radius of a sphere in the experimental frame of reference.

\begin{table}
	\centering
\begin{tabular}[c]{c|c|c|c}
	$L^\ast$ & $D_\parallel / D_0^S$ & $D_\perp / D_0^S$ & $D_r / D_r^S$ \\%
	\hline%
	0.02 (S) & 0.99 & 0.99 & 0.97 \\
	0.10 (C) & 0.97 & 0.95 & 0.85 \\
	0.24 (A) & 0.93 & 0.89 & 0.69 \\
	0.30 (B) & 0.91 & 0.87 & 0.63
\end{tabular}%

	\caption{Single particle diffusive properties used in the simulations, obtained by SHM calculations.}
	\label{tab:diffusion}
\end{table}

\begin{figure}[ht]
		\centering
		\includegraphics[width=\figurewidth]{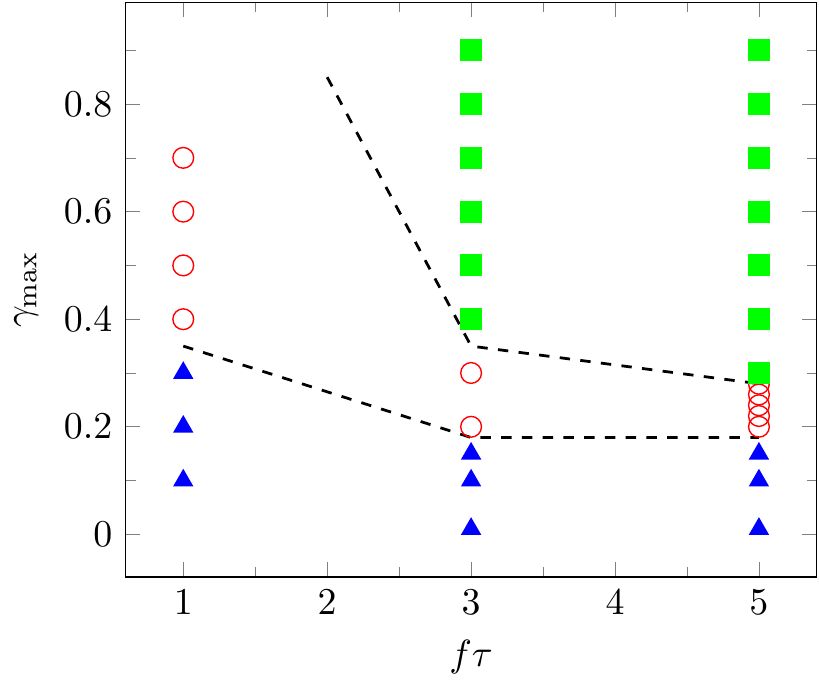}
		\caption{(Color online) Out-of-equilibrium states [I: twinned {FCC} (\textcolor{blue}{$\blacktriangle$}), II: disordered (\textcolor{red}{$\circ$}), III: sliding layers (\textcolor{green}{$\blacksquare$})] observed in the frequency-strain ($f\tau$-$\gamma_\mrm{max}$) plane of the parameter space in our BD simulations at state point~A, cf. Fig.\nobreakspace \ref {fig:phasediagram}. The dashed lines are tentative boundaries of the respective nonequilibrium states.}
		\label{fig:states_freq_strain}
	\end{figure}

	We simulate $(N=864)$ dumbbell particles subjected to Lees-Edwards~\cite{Lees1972} periodic boundary conditions.
	The systems are initialized in a crystalline state and run for $100 \tau$, at the frequencies $f=1\tau^{-1}$, $f = 3\tau^{-1}$ and $f=5\tau^{-1}$.
	The averages are calculated over $50$ and $250$ strain cycles in the steady state respectively.
Figure\nobreakspace \ref {fig:states_freq_strain} shows the parameters we have investigated in oscillatory shear conditions and the states we have identified.
For a summary of the equilibrium state points under consideration the reader is referred to Fig.\nobreakspace \ref {fig:phasediagram}.
The respective results are detailed in the Results section~\ref{section:results}.

\subsection{Trajectory analysis}

	\paragraph{Structure factors}
	The static structure factor $S(q_x, q_z)$ is evaluated in the velocity-vorticity plane at $q_y = 0$. 
	This reciprocal plane corresponds to a neutron experiment where the incident beam is parallel to the gradient direction of the shear flow.
	We calculate the structure factor directly from the {COM} coordinates of the dumbbells as
	\begin{align}
		S(q_x, q_z, 0) &= \left<\sum_{i=1}^{N} e^{-\imath \vec{q}\cdot\vec{R}_i}\right>
		\label{eq:soq_com},
	\end{align}
	where the angle brackets denote the trajectory average, which is taken in the steady state.

	\paragraph{Scattering intensity}
	The scattering intensity of the suspension $I(\vec{q})$ is the convolution of the {COM} structure and the distribution of scattering centers within each particle.
	The scattering amplitude $A(\vec{q}; \vec{u})$ describes the scattering of a single particle with orientation $\vec{u}$ and constant internal density at the scattering vector $\vec{q}$.
	Averaging the scattering amplitudes over all possible orientations yields the form factor $P(q)$.
	The total scattering intensity is calculated as 	
	\begin{align}
		I(\vec{q}) &= \left<\sum_{j,l} A(\vec{q}; \vec{u}_l) A(\vec{q}; \vec{u}_j) e^{-\imath \vec{q} \cdot (\vec{R}_l-\vec{R}_j)} \right>,
		\label{eq:ioq}
	\end{align}
	where the scattering amplitude of a homogeneous dumbbell tilted by the angle $\theta$ with respect to the scattering vector $\vec{q}$ is given by~\cite{Kaya2004}
	\begin{align}
		A(\vec{q}; \vec{u}) &= 4\pi R^3 \int\limits_{-L^\ast}^1 dt \cos\left(q \cos\theta R \left[t + L^\ast \right] \right).
		\label{eq:scatteramp}
	\end{align}

	\paragraph{Orientation}
	In order to analyze the orientational behavior of the particles, a measure of the orientation respecting the geometry of the system given is used. 
	We plot the mean orientation with respect to the Cartesian axes, which coincide with the velocity, shear gradient and vorticity axes, as 
	\begin{align}
		\left<P_2^\alpha\right>_\text{cycle}(t) &= \left<P_2(\cos\theta_\alpha)\right>_\text{cycle}(t),
		\label{eq:orientation_vscycle}
	\end{align}
	where $\alpha$ denotes the coordinates $x, y, z$ and $\theta_\alpha(t)$ the corresponding instantaneous angles.
	The average $\left< \right>_\text{cycle}$ denotes sampling all time steps with the imposed strain state $\gamma(t)$.

	The orientational correlations with respect to the time are investigated in terms of directional auto-correlation functions ({DACF}s) in the assumed steady state.
	Linearizing the equation of motion of the directors at some stationary state allows to define an effective rotational diffusion coefficient~\cite{Lenstra2001}.
	The {DACF}s $C_l(t) = \left<P_l(\vec{u}(t))\right>$, with $l$ being the order of the Legendre polynomial $P_l$, may be approximated by
	\begin{align}
		\label{eq:drot_dacf}%
		C_l(t) &= \exp\left(- l (l+1) \frac{t}{\tau_r}\right).
	\end{align}

	\paragraph{Center of mass order parameters}
	In order to distinguish crystalline structures, we use the local bond order analysis proposed by \cite{Steinhardt1983}.
	For each particle $i$ a vector $\vec{q}_l(i)$ is defined by the components
	\begin{align}
		\label{eq:q_lm}%
		q_{lm} &= \frac{1}{N_b(i)} \sum_{j=1}^{N_b(i)} Y_{lm}(\hat{\vec{R}}_{ij}),
	\end{align}
	where $Y_{lm}(\hat{\vec{R}}_{ij})$ are the spherical harmonics for the normalized separation vectors $\hat{\vec{R}}_{ij}$, and $N_b(i)$ is the number of the $i$-th particle's neighbors.

	\begin{align}
		\label{eq:ql}%
		\overline{q}_l(i) &= \frac{4\pi}{2l + 1}\sum_{m=-l}^{l}\left|{q}_{lm}(i)\right|^2,
	\end{align}
	\begin{align}
		\label{eq:wl}%
		\overline{w}_l(i) &= {\left(\sum_{m=-l}^{l}\left|{q}_{lm}(i)\right|^2\right)^{-3/2}} \\ \nonumber%
			&\times \sum\nolimits
			\begin{pmatrix}
				l & l & l \\
				m_1 & m_2 & m_3
			\end{pmatrix}%
			{q}_{{lm_1}}(i){q}_{{lm_2}}(i){q}_{lm_3}(i),
	\end{align}
	where the second sum runs over all $-l \le m_j \le l$ fulfilling ${m_1 + m_2 + m_3 = 0}$.

	We define the ensemble averaged order parameters as
	\begin{align}
		\label{eq:OPQ_l}%
		\left<Q_l\right> &= \left<\frac{1}{N}\sum_{i = 1}^{N} \overline{q}_l(i)\right>, \ \text{and}\\
		\label{eq:OPW_l}%
		\left<W_l\right> &= \left<\frac{1}{N}\sum_{i = 1}^{N} \overline{w}_l(i)\right>,
	\end{align}
	where the angle brackets $\left< \dots\right>$ denote the time average in the steady state.
	In this work the {COM} order is monitored in terms of averaged local order parameters $\left<Q_4\right>$ and $\left<W_4\right>$. 
	These order parameters are sensitive to the configurations of the neighborhoods of solid-like particles. 
	In particular, {FCC} and hexagonal close packed ({HCP}) structures are separated by a change of the sign of $\left<W_4\right>$.

	\paragraph{Radial distribution function and Enskog collision rate}
	The radial distribution function ({RDF}) of the beads is defined as 
	\begin{align}
		g(\vec{r}) &= \frac{1}{N \rho}\left<\sum_{i}\sum_{j}\delta\left(\vec{r} -(\vec{r}_i - \vec{r}_j)\right)\right>.
		\label{eq:rdf}
	\end{align}
	In a homogeneous and isotropic system the {RDF} depends only on the distance $r = \left|\vec{r}\right|$.

	From a structural point of view, it is interesting to investigate the collision probability (or rate) during the shear and the phase transitions. 
	It gives a rough picture of the average configurational freedom of the dumbbells under shear. 
	The Enskog collision rate for hard spheres with diameter $\sigma$ is given by~\cite{HansenMcDonald} 
	\begin{align}
		\Gamma_E &= g(\sigma) \Gamma_0 ,
		\label{eq:enskog_collision}
	\end{align}
	where $g(\sigma)$ is the contact value of the {RDF}, and $\Gamma_0$ is the collision frequency in the dilute gas which is the ratio of the mean velocity and the free path in a suspension of hard spheres.
	In the Enskog approximation, which neglects correlated collisions, the self-diffusion coefficient is inverse proportional to the contact value
	\begin{align}
		D_E &= \frac{3 k_B T}{2 m \Gamma_E},
		\label{eq:enskog_diffusion}
	\end{align}
	for a particle of mass $m$.
	Since our particles have smooth potentials, the {RDF} does not have its first maximum exactly at $r=\sigma$.
	Therefore, we take the height of the first maximum $g(r_\mrm{max})$ as a measure for the collision rate.

\section{Results and Discussion}
\label{section:results}

\subsection{Steady-state structures and transitions}

\begin{figure*}[ht]
		\includegraphics[width=\linewidth]{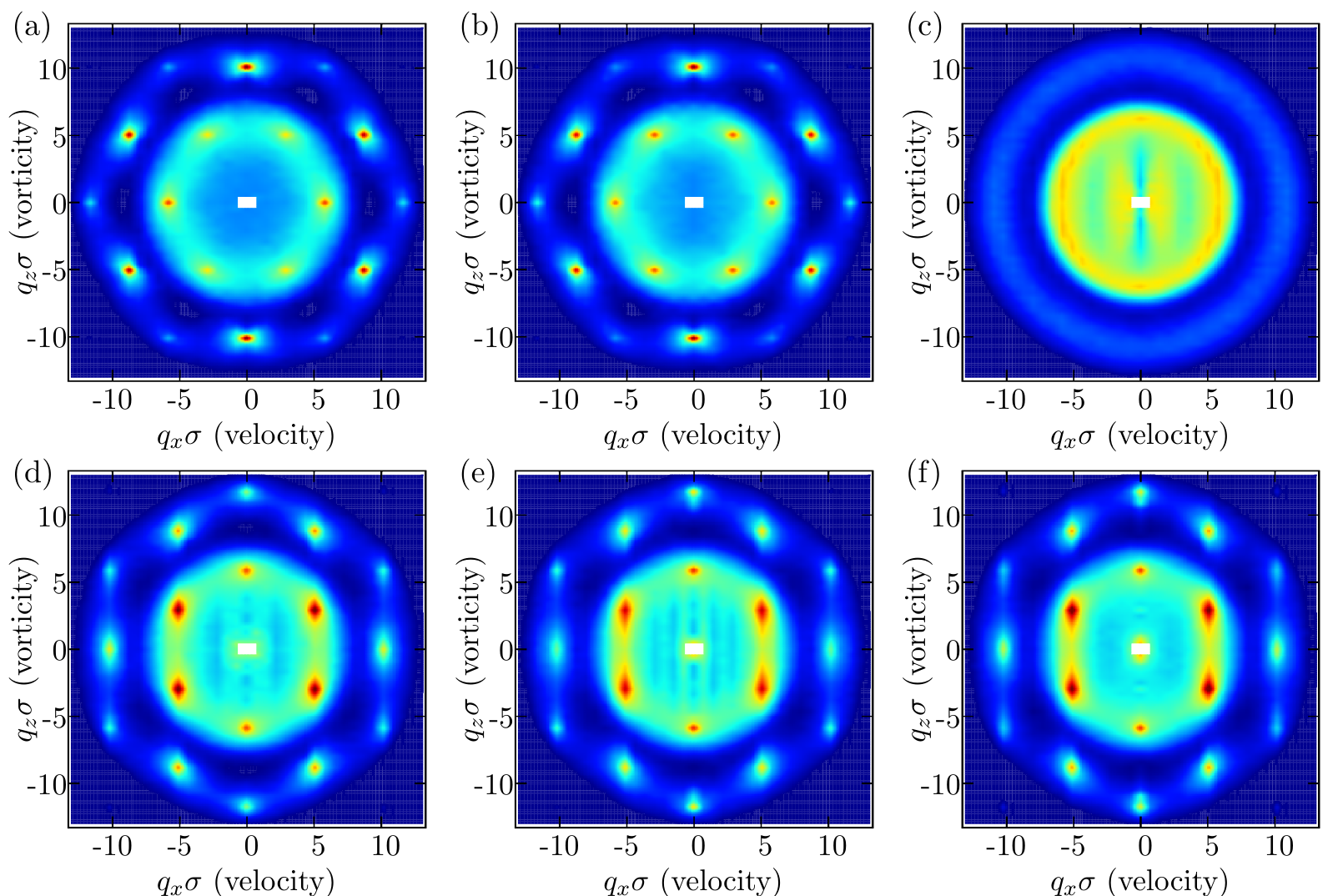}
		\caption{(Color online) Scattering intensities $I(\vec{q})$ ($250$ cycles averaged) in the $q_y = 0$ plane of sheared dumbbell suspensions ($L^\ast = 0.30, \phi = 0.55$, state point B) at frequency $f = 5\tau^{-1}$. From top left to bottom right: increasing strain amplitude $\gamma_\text{max}$: $0.05$ (a), $0.1$ (b), $0.2$ (c), $0.3$ (d), $0.5$ (e) $1.0$ (f)}
		\label{fig:ioq_freq5.0_lstar0.30}
	\end{figure*}

We start with the discussion of BD simulated scattering intensities for selected state points.
In Fig.\nobreakspace \ref {fig:ioq_freq5.0_lstar0.30} the scattering intensities $I(\vec{q})$ including the scattering amplitude calculated as defined in equations Eq.\nobreakspace \textup {(\ref {eq:ioq})} and Eq.\nobreakspace \textup {(\ref {eq:scatteramp})} are shown in the velocity-vorticity plane.
The data corresponds to state point~B, cf. Fig.\nobreakspace \ref {fig:phasediagram}, at a frequency of $5\tau^{-1}$ for selected strains.

The intensity plots illustrate the transition from the low strain twinned crystal state (I) to the high strain sliding layer state (III), corroborating with earlier results on a smaller aspect ratio~\cite{Chu2015}. 
In all ordered states a hexagonally ordered plane is parallel to the velocity-vorticity plane in the shear flow framework.
A common feature of the scattering plots is the fact that the maxima may be seen on rings of constant scattering vector magnitude.
At low strain amplitudes (Fig.\nobreakspace \ref {fig:ioq_freq5.0_lstar0.30}a,b) a shear twinned {FCC} crystal is stable, which has a densely packed direction perpendicular to the velocity direction, while at high strains (Fig.\nobreakspace \ref {fig:ioq_freq5.0_lstar0.30}d-f) a dense direction parallel to the velocity is favorable.
Here, the fully ordered systems tend to form two-dimensional {HCP} layers in the velocity-vorticity plane.
These planes are the most densely packed crystallographic planes for {FCC} and {HCP} crystal structures and their stacking sequence determines the crystallographic type. 
	
Similar transitions had been observed in spherical systems~\cite{Ackerson1988, Loose1994, Besseling2012}.
However, slightly anisotropic dumbbells introduce an additional, orientational degree of freedom, and, although weak, it  leads to a more abrupt transition compared to the spherical case, as we have previously demonstrated for a smaller aspect ratio~\cite{Chu2015}.
In equilibrium terms it could be said the mild anisotropy qualitatively change the transition from being continuous to discontinuous.
This phenomenon is already present in the series of scattering intensities, as we observe a fully molten state for intermediate states, cf.~Fig.\nobreakspace \ref {fig:ioq_freq5.0_lstar0.30}c.
In contrast, for the hard spherical reference case, we always detect order in form of crystalline hybrids~\cite{Chu2015}.
At a strain amplitude about $\gamma_\mrm{max} \approx 0.2$ the system does not show long-range correlation and is nearly isotropic (Fig.\nobreakspace \ref {fig:ioq_freq5.0_lstar0.30}c).
Close to this transition an anisotropy may be observed at very small scattering vectors in the reciprocal velocity-vorticity plane.
Above this isotropic state hexagonal layers perform a zig-zag trajectory, which is indicated by a $30 \,{}^\circ$ tilt of the scattering pattern.
In the present case we conclude from comparison to the idealized picture and from the reasoning concerning the volume fractions given by~\citet{Ackerson1990a} that the {COM} motion of the dumbbells follow strongly registered trajectories while maintaining two-dimensional hexagonal in-plane order.

\begin{figure}[ht]
		\centering
		\includegraphics[width=\figurewidth]{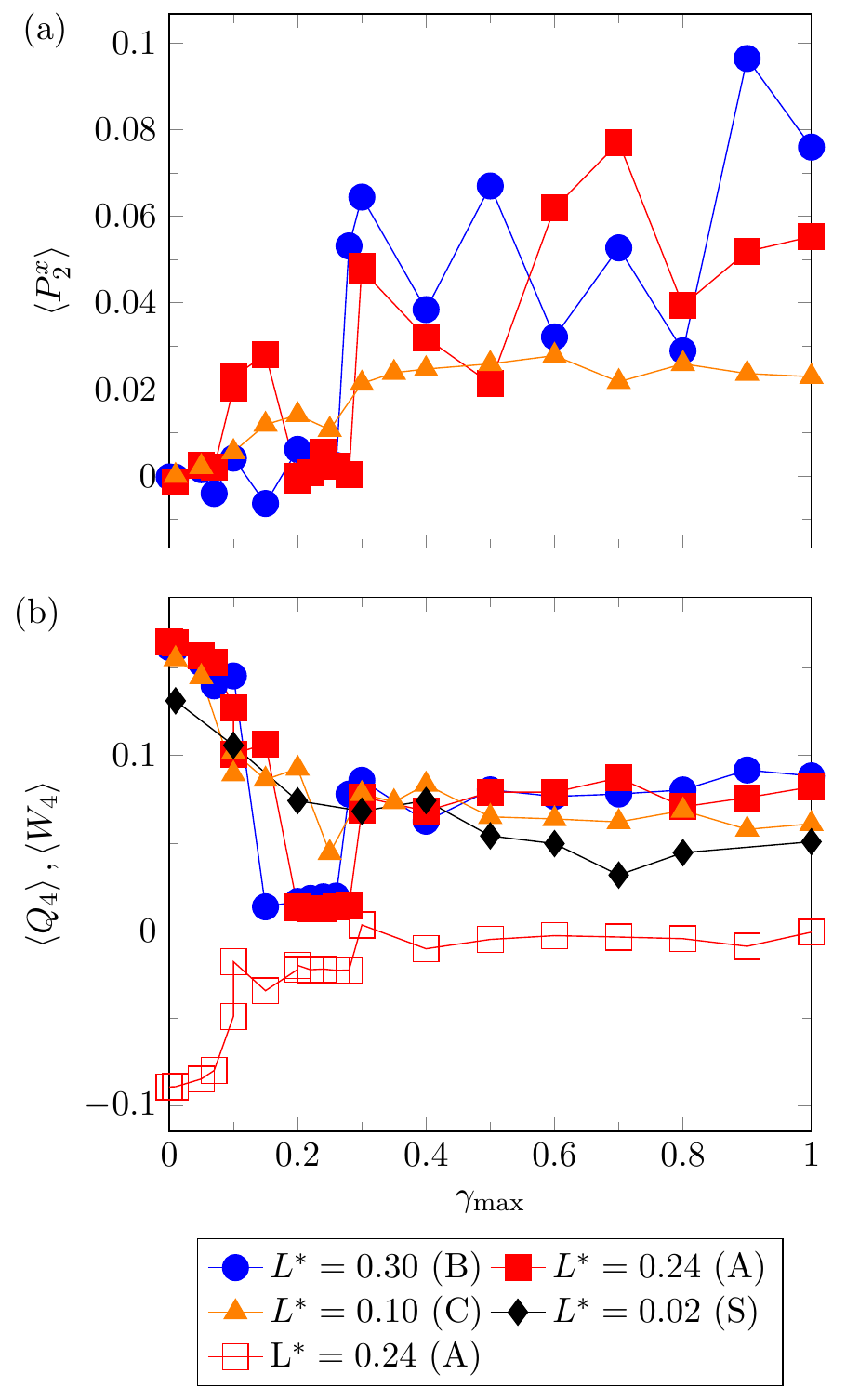}
		\caption{(Color online) Order parameters for state points A and B at $f = 5\tau^{-1}$.
(a) {Averaged orientational order parameter in flow direction $\left<P_2^x\right>$, and}
(b) {translational order characterized by $\left<Q_4\right>$ (filled symbols) and $\left<W_4\right>$ (open squares).}%
		}
		\label{fig:average_op_freq5.0}
	\end{figure}

In the following, we investigate the particles' orientational and translational structure under the action of shear.
Figure\nobreakspace \ref {fig:average_op_freq5.0} summarizes the structural information in terms of averaged order parameters for state points A and B, cf. Fig.\nobreakspace \ref {fig:phasediagram}, at a frequency of $5\tau^{-1}$.
Figure\nobreakspace \ref {fig:average_op_freq5.0}a shows that the average $\left<P_2^x\right>$, quantifying the average dumbbell orientation along the flow direction, is slightly increasing with strain for both aspect ratios in the {FCC} state until it drops to zero in the disordered state and is non-zero again in the high-strain regime.
In Fig.\nobreakspace \ref {fig:average_op_freq5.0}b the translations order, described by $\left<Q_4\right>$, is shown for the dumbbells compared to the nearly hard sphere reference case~(S). 
Thus, the non-monotonicity of the orientational measure with respect to the strain amplitude is due to the loss of long-range order.
The latter is not preserved in the transition region for sufficiently high aspect ratios, at which neighboring layers of different orientation cannot pass smoothly any more.
The behavior at the transition shows clearly that the orientational and translational order changes abruptly in a discontinuous fashion for dumbbells, while, in contrast, the transition in suspension of hard \emph{spheres} has a continuous character.
This behavior is retained for dumbbells with a slender elongation of $L^\ast = 0.10$ (system~C), which show a ordered state at all investigated strain amplitudes.
The average $\left<P_2^x\right>$ for state point~C shows a slight increase at low strain amplitudes as well approaching a plateau value of about $0.02$ in a rather monotonous fashion beyond $\gamma_\mrm{max} \gtrsim 0.3$.
This corresponds well to our previous finding that long-time stress correlations due become important for dumbbells above approximately $L^\ast =0.15$~\cite{Heptner2015}.
Additionally, the negative $\left<W_4\right>$ in state~I indicates {FCC} dominated structure, in state~III this parameter is vanishes on average indicating loss of {FCC} order.

Moreover, the steady state structures and transitions depend on the shear frequency, which is summarized in the nonequilibrium phase diagram in  Fig.\nobreakspace \ref {fig:states_freq_strain}.
The diagram shows the phases depicted for various strain amplitudes $\gamma_\mrm{max}$ versus frequency $f$ for state point~A.
In our simulations, state~III does not appear at $f = 1/\tau$ whereas state~I is stable up to $\gamma_\mrm{max} \approx 0.30$.
For higher frequencies $f=3/\tau$ and $f=5/\tau$ we observe the ordered state~III above an amplitude of approximately $0.3$, moving slightly to lower strains on increasing frequencies, which is sketched by the dashed lines in Fig.\nobreakspace \ref {fig:states_freq_strain}.
	
\begin{figure*}[ht]
		\centering%
		\includegraphics[]{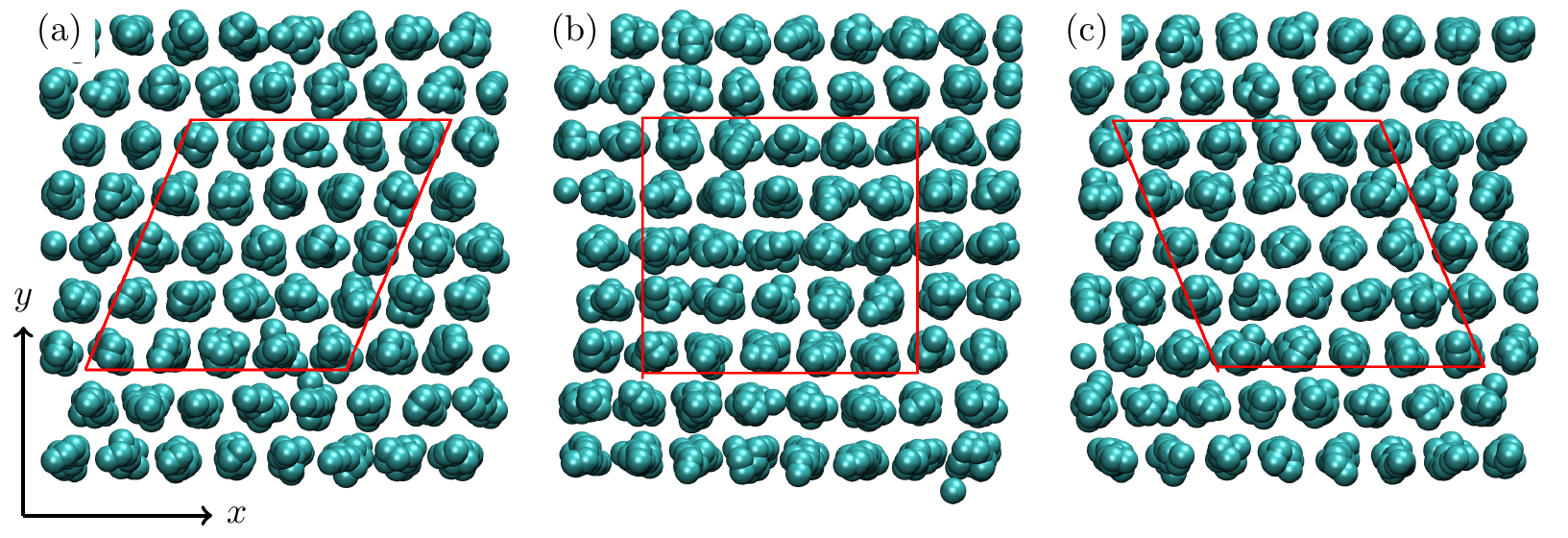}
		\caption{(Color online) Simulation snapshots at prominent points in the shear cycle for $\gamma_{{\text{max}}} = 0.3, f = 1\tau^{-1}$ for state point~A in the twinned FCC state~I at (a) maximum, (b) zero, and (c) minimum instantaneous strain in the shear cycle. The particle radii are scaled by $1/2$ for a better view of the structure.}
		\label{fig:snapshots_twin}
	\end{figure*}

\begin{figure*}[ht]
		\centering%
		\includegraphics[]{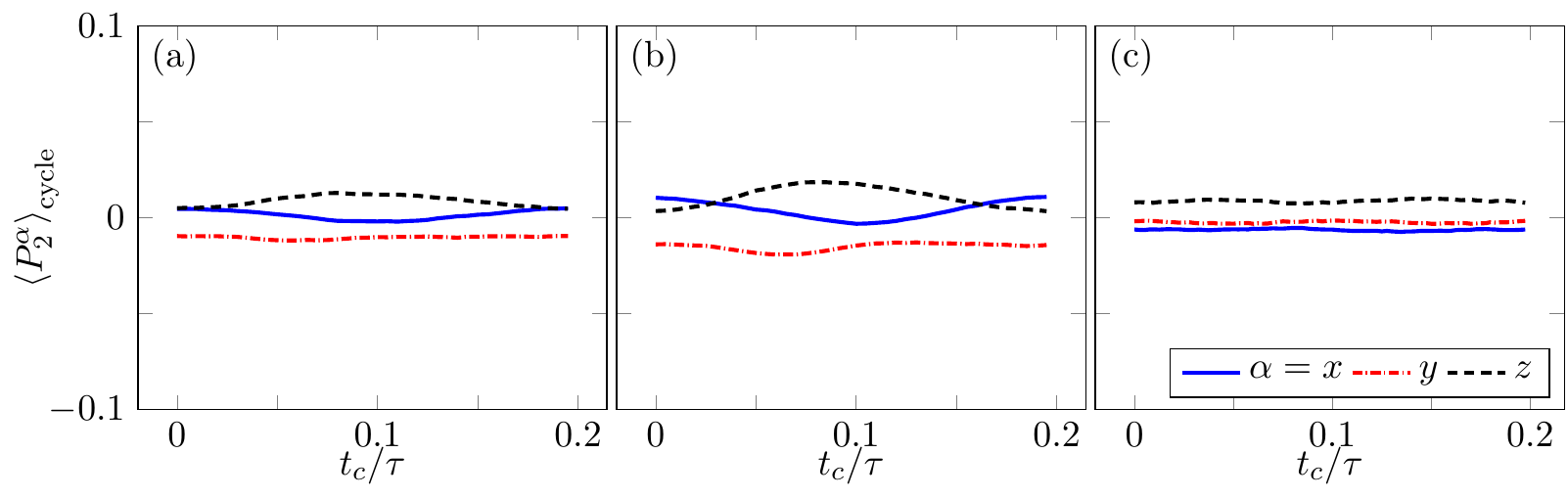}
		\caption{(Color online) Time-resolved orientation within one cycle (averaged over 250 cycles) at frequency $f=5\tau^{-1}$, elongation $L^\ast = 0.30$ (state point B),  volume fraction $\phi = 0.55$, and for different strain amplitudes: (a) $\gamma_\mathrm{max} = 0.05$, (b) $\gamma_\mathrm{max} = 0.10$, and (c) $\gamma_\mathrm{max} = 0.15$. }
		\label{fig:orientation_vstime_twinnedfcc}
	\end{figure*}

Let us now characterize the respective nonequilibrium states in more structural detail.\\
	{\it Twinned-FCC regime (I)}:
	At low strain amplitudes a shear-twinned {FCC} dominated structure is observed in the steady state.
	Figure\nobreakspace \ref {fig:snapshots_twin} shows a series of snapshots taken at distinctive points in the strain cycle at which the instantaneous strain is minimal ($\gamma(t)=-\gamma_\mathrm{max}$), zero ($\gamma(t) = 0$) or maximal respectively.
	In this state, particles may follow the oscillatory flow and transfer between the triangular voids offered by neighboring layers.
	These void spaces are accessible in the vicinity of the extrema of the strain cycle, in between the particles are forced to pass particles of adjacent layers closely, which is referred to as bridge-stacking.
	This behavior finds expression in the transient orientation which is subtly coupled to the strain cycle through particle interaction.
	At these low strain amplitudes, $\gamma_\mathrm{max} < 0.1$, the orientation shows a subtle interplay between the velocity and the vorticity axes while the amplitude in both is very small. This is exemplified in Fig.\nobreakspace \ref {fig:orientation_vstime_twinnedfcc}a, where we show the time-resolved orientation in one shear cycle. 
	In approaching the transition to the high strain state, $0.1 \le \gamma_\mathrm{max} \le 0.2$, a slight separation of the velocity axis is observed, cf.~Fig.\nobreakspace \ref {fig:orientation_vstime_twinnedfcc}c, while the amplitudes of the $\left<P_2^x\right>_\mrm{cycle}$ and $\left<P_2^z\right>_\mrm{cycle}$ are still in the order of the average.
	The shear twinned {FCC} has a signature in the structure factor, see Fig.\nobreakspace \ref {fig:ioq_freq5.0_lstar0.30}a,b: the inner peaks are forbidden for equilibrium {FCC} crystals, in the present case we clearly observe non-vanishing peaks on the first ring and we see that their magnitude grows on increasing strain.
	Additionally, let it be noted, that the peaks on the velocity axis ($q_z=0$) are the first to rise at very low strain amplitudes.
	Although the shear twin is not fully developed at strains less than $0.2$, the scattering intensity reveals that the equilibrium crystal is disturbed sufficiently from $\gamma_\mrm{max} \gtrsim 0.05$.
	Let it be noted that the present very dense systems are crystalline in equilibrium and it has been shown that the crystalline state becomes stable for lower amplitudes on increasing density~\cite{Ackerson1990a}.
	\\
	
\begin{figure}[ht]
		\centering%
		\includegraphics[width=\figurewidth]{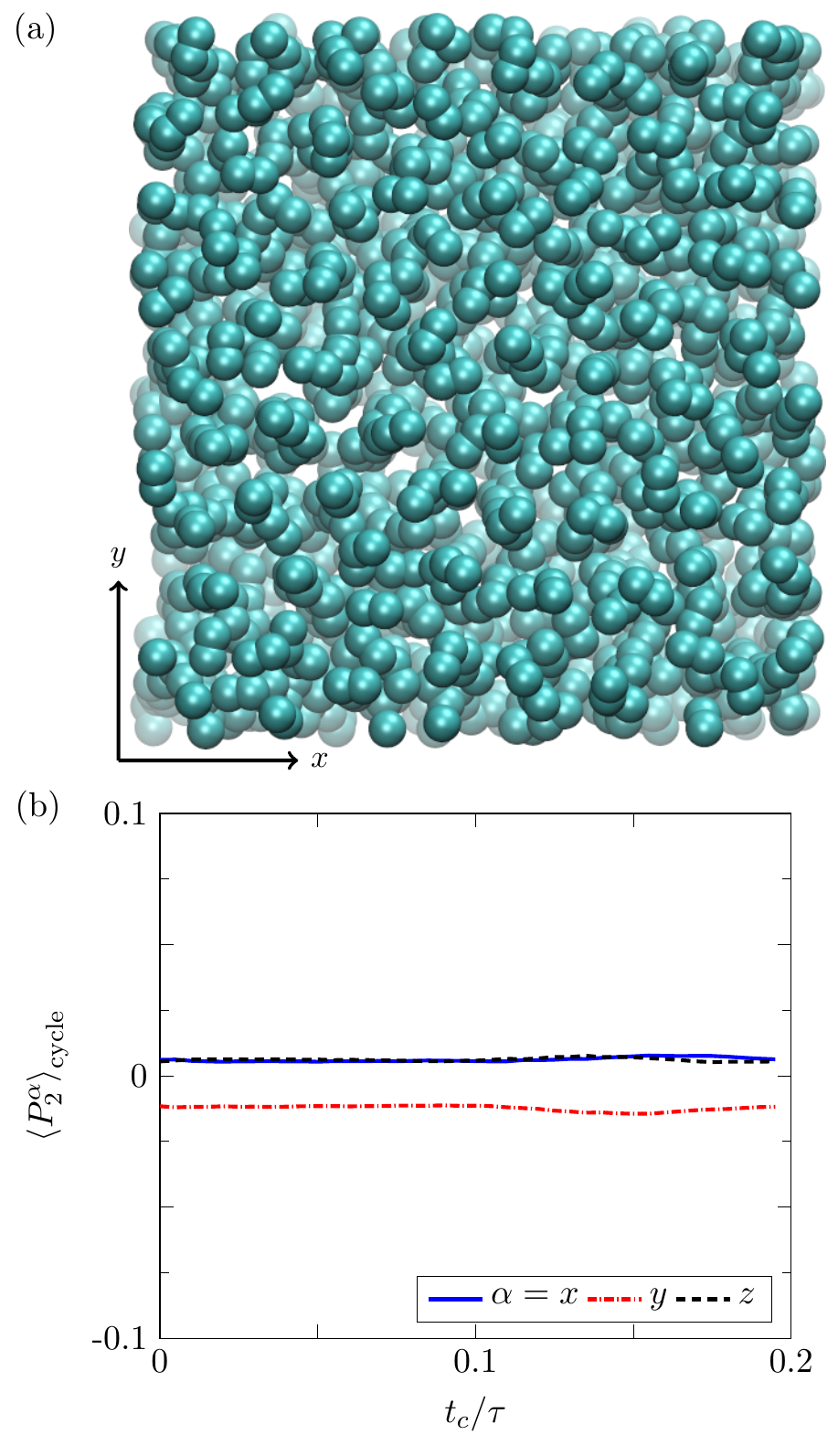}
		\caption{(Color online) System (B) in the fully disordered state II at $\gamma_\mathrm{max} = 0.2$, snapshot (a) and cycle averaged orientational order parameters $\left<P_2^\alpha\right>_\mrm{cycle}(t_c)$ (b).}
		\label{fig:state_ii}
	\end{figure}

	{\it Intermediate disordered state (II)}:
	While for suspensions of hard spheres a disordered state in between the low-strain twinned-{FCC}~(I) and high-strain sliding layer~(III) regimes is not observed, in fact, we find stable hybrid structures on reduction of the anisotropy, for sufficiently elongated dumbbells the low-strain structure always melts fully at intermediate strain amplitudes. 
	We find neither long-ranged translational order nor any orientational correlations whatsoever, which is confirmed considering the scattering intensities in Fig.\nobreakspace \ref {fig:ioq_freq5.0_lstar0.30}c.
	A representative snapshot and cycle-averaged orientations are shown in Fig.\nobreakspace \ref {fig:state_ii}, clearly demonstrating disorder.
  Evidently, this is a distinctive behavior introduced by the orientational degree of freedom of the particles with a sufficient elongation.\\
      
\begin{figure}
		\centering
		\includegraphics[width=\figurewidth]{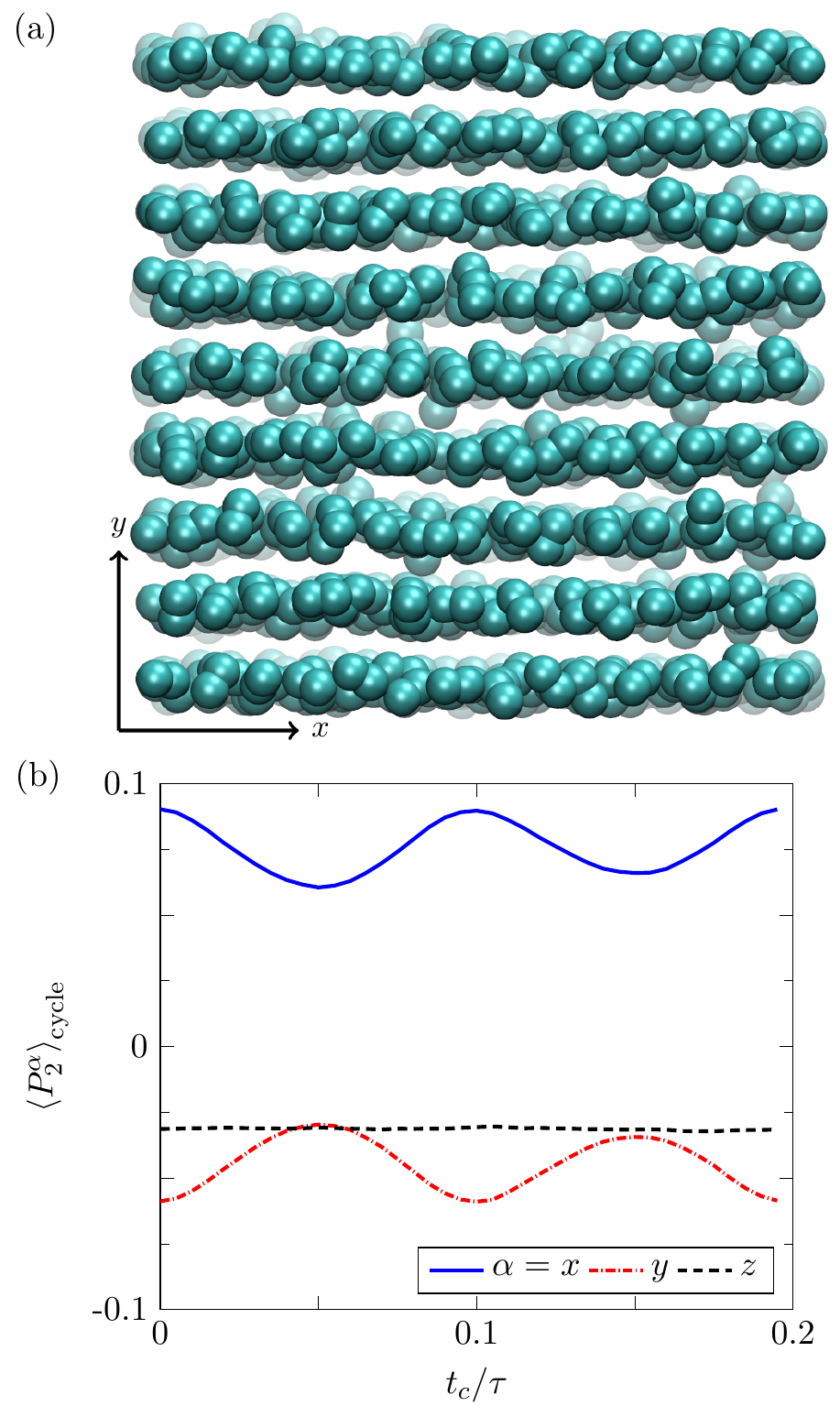}
		\caption[]{(Color online) Snapshot in the flow-gradient ($x-y$) plane (a) and oriental order parameters (b) in velocity ($x$), gradient ($y$) and vorticity ($z$) directions versus strain cycle at $\gamma_{\text{max}} = 1.00$ for state point~B in the high shear state~III, where the orientations exhibit a finite order modulated by the imposed shear.}
		\label{fig:orientation_vstime2_strain1.00}
	\end{figure}

	{\it Sliding layer regime (III)}:
	While the centers of masses perform a zig-zag motion, a tendency of the particles' orientation towards the velocity axis is observed, directly after reaching the critical strain to assemble into velocity oriented layers. 
	Figure\nobreakspace \ref {fig:orientation_vstime2_strain1.00} shows the cycle averaged orientation in the high strain regime, where a dense direction of each layer is aligned with the velocity direction.
	Here, the velocity and gradient directions of the directors are clearly modulated by the shear cycle, while the vorticity direction is essentially flat.
	In this case, the modulus of the $P_2^x$ cycle average is maximal at times, when the instantaneous strain vanishes ($\gamma(t) = 0$), and its amplitude is about $0.01$ and the average $0.06$.
	On average the directors slightly tend to be parallel to the velocity and perpendicular to the gradient.
	This coincides with a decoupling of the orientation from the imposed strain $\gamma(t)$, where the amplitudes of the $\left<P_2^\alpha\right>_\mrm{cycle}$ are significantly smaller than their respective averages.
	
\subsection{Kinetic properties}

\begin{figure}[ht]
		\centering
		\includegraphics[width=\figurewidth]{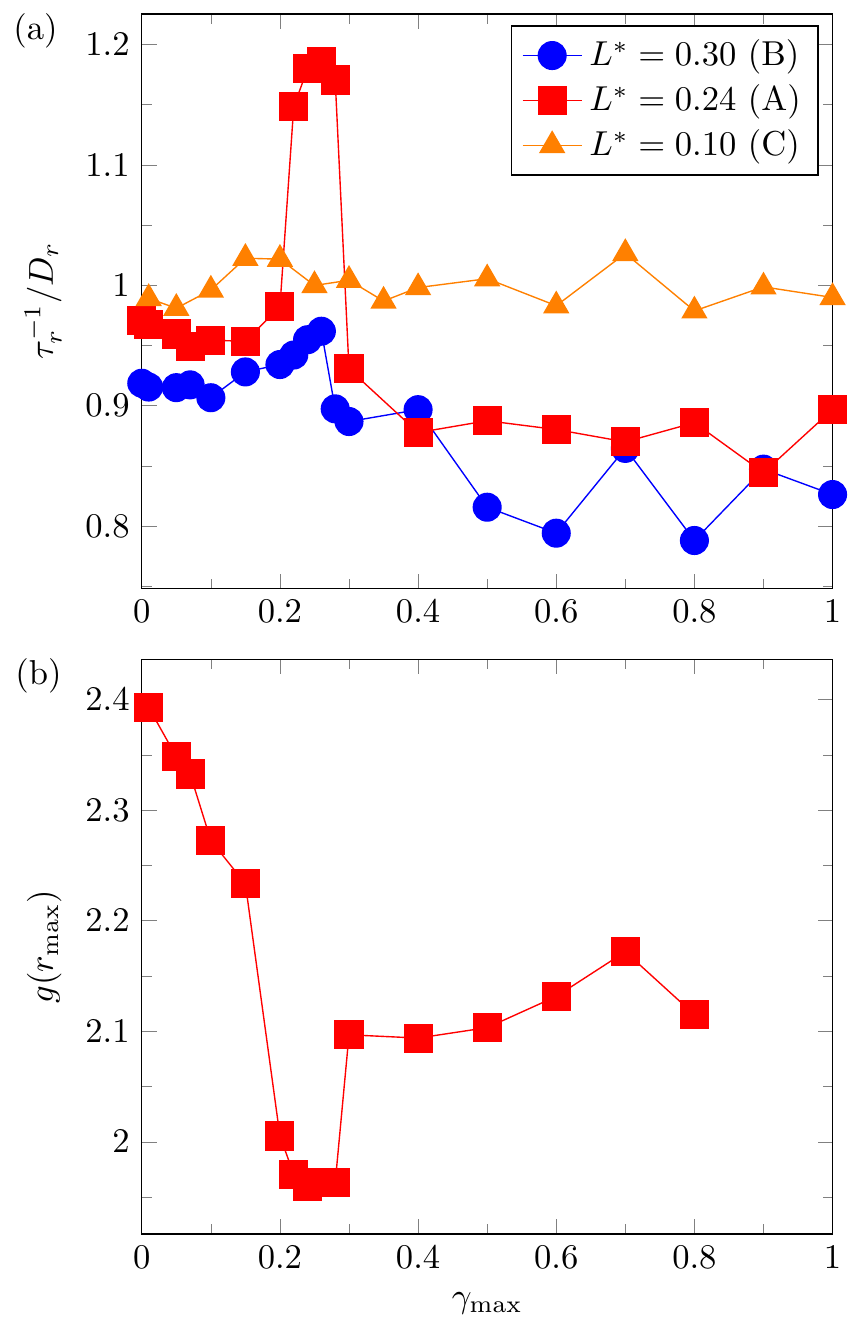}
		\caption{(Color online) %
			(a) Orientational relaxation constants from the decays of the $C_1$ correlation functions for state points~A, B and C, and 
			(b) contact values of the radial distribution function $g(r)$ on increasing strain amplitude for state point~A.
		}
		\label{fig:drot_dacf}
	\end{figure}

	Let us now turn our focus to rotational relaxation properties and particle collision rates. 
	Figure\nobreakspace \ref {fig:drot_dacf}a shows the inverse relaxation constant $\tau_r^{-1}$ obtained from the exponential decays of the orientational autocorrelation functions according to Eq.\nobreakspace \textup {(\ref {eq:drot_dacf})} normalized by their short-time values at infinite dilution $D_r$ in absence of any external field.
	The effective rotational decay is enhanced in the disordered state while it is similar to the equilibrium case in the low strain regime and a significant slowing down is observed in the high strain state.
	At small strain amplitudes the values are slightly smaller than unity as we expect from the analysis of the equilibrium behavior with respect to volume fraction and elongation~\cite{Heptner2015}.
	In the regularly structured twinned {FCC} regime (I) the diffusion is basically constant with increasing strain amplitude.
	In the transition region $0.2 < \gamma_\mathrm{max} < 0.3$ we observe an elevated orientational diffusion where it is steeply curved with respect to the strain amplitude.
	On entering the high strain regime it jumps back to a value close to its initial value at rest.
	With increasing strain amplitude the orientational diffusion then slightly decreases.
	In the high strain regime we observe a state showing enhanced coupling of the dumbbells' orientations in space and time.
	For the higher aspect ratio (system~B) we observe similar behavior, while the peak in the disordered state (II) at about $\gamma_\mrm{max} \approx 0.25$ is much less pronounced.
	Also, on increasing the elongation, the normalized inverse time scales are smaller than in the former system~(A).
	This can be explained considering the packing effect from our previous study~\cite{Heptner2015}, where we show that packing gets important for long-time relaxation from $L^\ast\gtrsim 0.3$.
	At the low aspect ratio of state point~C, which does not show a fully disordered state, we observe essentially unhindered rotational relaxation and virtually no influence of the shear amplitude.

	The transition behavior of the rotational diffusion corresponds well to the contact value analysis, allowing us to connect structure and kinetics.
	Figure\nobreakspace \ref {fig:drot_dacf}b shows the values of the {RDF} $g(r)$ at contact $(r=r_\mrm{max})$ of system~A with $L^\ast = 0.24, \phi = 0.55$.
	On approaching the melting strain at about $\gamma_\mathrm{max} = 0.2$ the contact value $g(\sigma) = \Gamma_E / \Gamma_0$ rises slightly.
	At the transition from the ordered shear twinned system~(I) to the disordered state~(II) the contact value shows a distinct jump of about $10 \,\%$.
	A smaller jump is observed at the transition to the ordered high strain regime where the contact value is less than in in the disordered state and reaches a plateau at about $g(r_\mrm{max}) \approx 2.1$.
	Following the inverse relationship of diffusion and contact value in the Enskog approximation Eq.\nobreakspace \textup {(\ref {eq:enskog_diffusion})}, thus the diffusivity increases drastically in the transition region~(II).

	\subsection{Comparison to experiments}
	
	\begin{figure}[ht]
		\centering
		\includegraphics[width=\figurewidth]{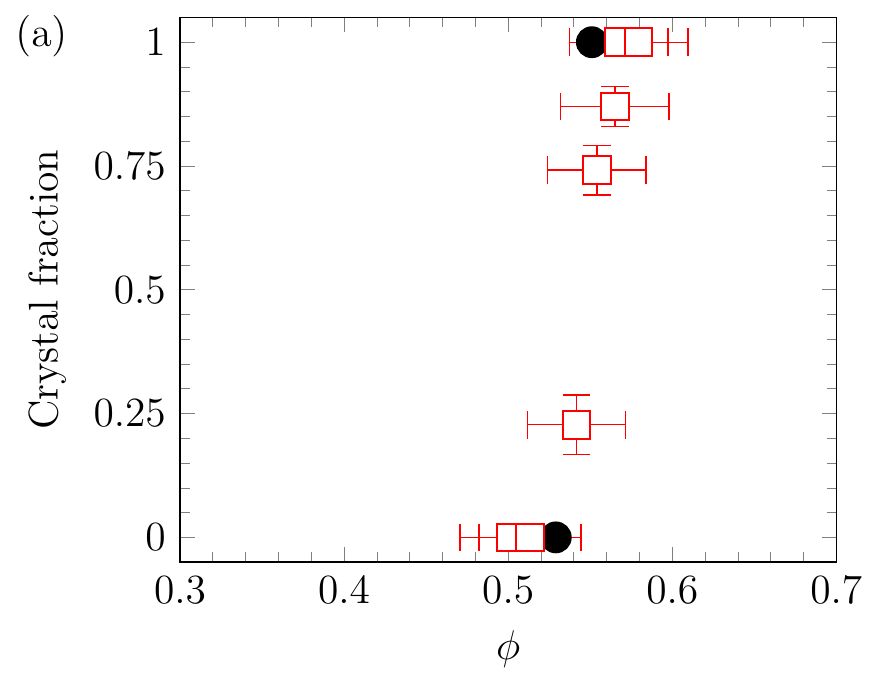}
		\includegraphics[width=\figurewidth]{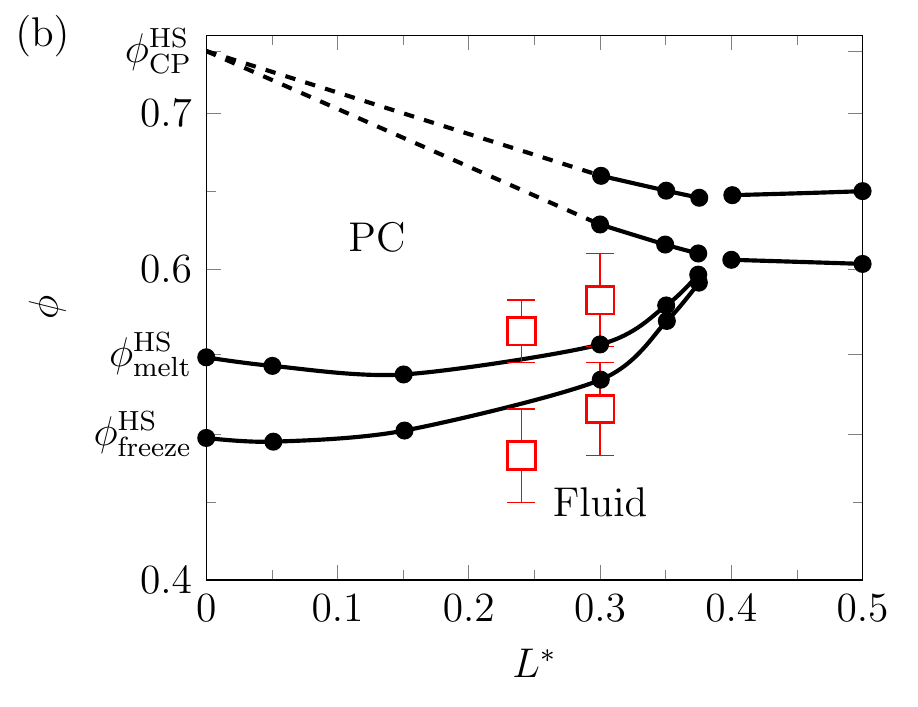}
		\caption{(Color online) %
			Phase equilibrium (a) of hard dumbbells at $L^\ast \approx 0.3$ from crystallisation experiments. The experimental phase diagram (b) (denoted by red squares ($\textcolor{red}{\square}$)) is compared with the prediction of {MC} simulations (solid black line ($\textcolor{black}{\bullet}$)~\cite{Marechal2008}) for $L^\ast = 0.24$ and $L^\ast = 0.30$.
		}
		\label{fig:experimental_equilibria}
	\end{figure}

	\begin{figure*}[ht]
		\centering
		\includegraphics[width=2\figurewidth]{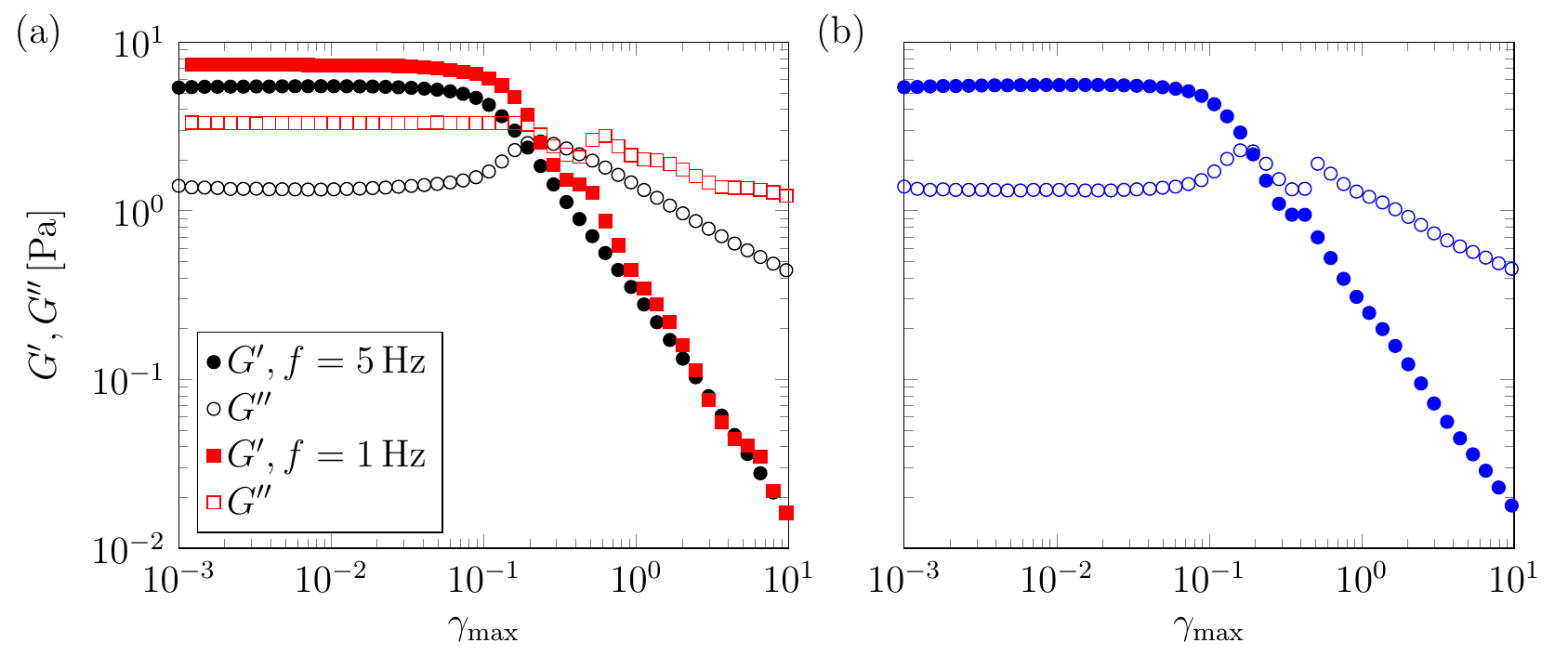}
		\caption{(Color online) %
			(a) Dependence of $G^\prime$ and $G^{\prime\prime}$ on increasing $\gamma_\mathrm{max}$ for hard dumbbells with $L^* \approx 0.30$ in the plastic crystal phase under oscillatory shear at $f = 1 \,\mathrm{Hz}$ (circles) and $f = 5 \,\mathrm{Hz}$ (squares). This measurement is performed with the default setting ($100 \,\mathrm{s/point}$).
			(b) Shear moduli versus strain amplitude $\gamma_\mathrm{max}$ for the same system phase that is measured under oscillatory shear of $f = 1 \,\mathrm{Hz}$ with $500 \,\mathrm{s/point}$. 
			The filled symbols denote $G'$, while open symbols represent $G''$.
		}
		\label{fig:experimental_rheology}
	\end{figure*}
	
	In the following we introduce novel experiments on the rheology for the 'colloidal nitrogen' case at $L^\ast = 0.3$~(B).
	The details on the experimental and synthesis procedures may be obtained from our previous papers~\cite{Chu2012,Chu2015}.
	Comparison to the prediction of the phase diagram in~Fig.\nobreakspace \ref {fig:experimental_equilibria}b shows that the coexistence region (fluid/{PC}) shrinks on increasing elongation and the phase boundaries of the experimental systems fit very well.
	A glance at~Fig.\nobreakspace \ref {fig:experimental_equilibria}a also confirms that the fraction of crystalline sample in biphasic region is linear in concentration, and, thus, the equilibrium properties of the experimental correspondent of system~B are well-defined.
	The yielding behavior of the hard dumbbells with $L^\ast \approx 0.30$ in the plastic crystalline phase is displayed in Fig.\nobreakspace \ref {fig:experimental_rheology}a. 
	The experimental data indicate kinetic differences as the rheology of the system~A is retained by either the frequency or the number of shear cycles per point:
	While the yielding behavior of hard dumbbells with $L^\ast \approx 0.30$ in the plastic crystalline phase displayed in Fig.\nobreakspace \ref {fig:experimental_rheology}a exhibits one event under oscillatory shear at $f = 1 \,\mathrm{Hz}$ there are two yielding events at $f = 5 \,\mathrm{Hz}$.
	Compared with the former frequency, the number of applied shear cycles is increased by five times within the same measurement time for the experiment at $f = 5 \,\mathrm{Hz}$.
	It is necessary to mention that the double yielding event is observed in the oscillatory shear field with $f = 1 \,\mathrm{Hz}$ as well when the measurement time is prolonged by five times as shown in Fig.\nobreakspace \ref {fig:experimental_rheology}b.
	Based on these three sets of experiments, it is concluded that the hard dumbbells with $L^* \approx 0.30$ in the plastic crystalline phase can show the same double yielding behavior as the hard dumbbells with $L^* \approx 0.24$, but the former needs more or faster oscillations to achieve the steady state. 
	The system of hard dumbbells with $L^* \approx 0.30$~(B) is closer to the glassy state predicted by~\citeauthor{Zhang2009} than that with $L^* \approx 0.24$~(A) at the same volume fraction of $0.6$.
	Due to the expected slowdown of the dynamics in the vicinity of the glass transition in the former case, stronger and longer oscillations are required to induce the same structural change as that with $L^* \approx 0.30$.

	As detailed above, in the {BD} simulations at constant frequency ($f=5\tau^{-1}$) (Fig.\nobreakspace \ref {fig:average_op_freq5.0}) we observe the transition at nearly the same strain amplitude for systems~A and~B, while for both cases this transition clearly occurs at lower amplitudes.
	Moreover, rheo-SANS experiments have been carried out to investigate the underlying shear-induced structural evolution that corresponds to the double yielding behavior.
	For a direct compilation of scattering data and rheology see~Fig.\nobreakspace \ref {fig:fig_s1} in the appendix. 
	The rheo-SANS experiments clearly demonstrate that the longer dumbbells~(B) undergo the same structural evolution to those with $L^* \approx 0.24$~(A), corresponding to the simulation results.
	With the increasing applied shear strains, fully crystallized longer dumbbells undergo the phase transition from twinned {FCC}~(I) to the partially orientated sliding layer state~(III), while being disordered in between~(II).
	The main difference in the experimental realization is that the longer dumbbells need larger P\'{e}clet numbers $\mrm{Pe}_r$ to induce the same nonequilibrium states as compared with hard dumbbells with $L^* \approx 0.24$, which we summarize in~Table\nobreakspace \ref {tab:peclet_numbers}.
	It should be noted that the simulated frequencies and, thus, the P\'eclet numbers are approximately five times as high as the experimental parameters.
	In our simulations, the transition from II to III, occurs nearly at the same values $\mrm{Pe}_r=4.56$ and $4.64$ for systems~A and B, respectively.

	\begin{table*}
	\centering
		\includegraphics[]{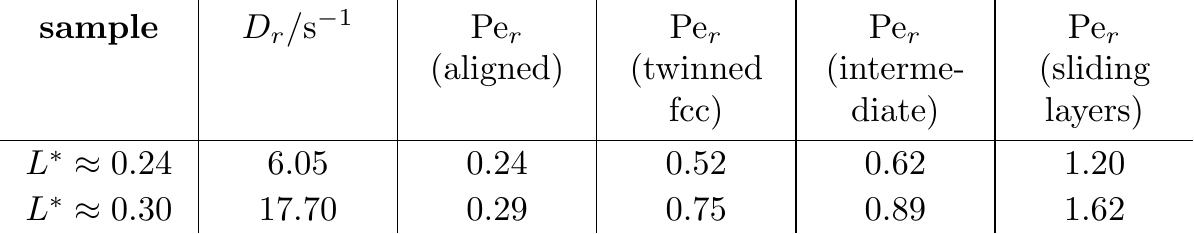}
		\caption{Comparison of critical $\mrm{Pe}_r$ at the formation of vorticity alignment structure, the twinned {FCC}, the intermediate structure and the partially oriented sliding layers in the hard dumbbells with $L^* \approx 0.24$ and $L^* \approx 0.30$. $D_r$ is obtained from DLS and DDLS measurements.}
		\label{tab:peclet_numbers}
	\end{table*}

\section{Conclusions}

	Following up on our previous study~\cite{Chu2015}, we have here provided more evidence that the mild anisotropy of dumbbell-shaped particles leads to qualitative changes in the nature of the nonequilibrium phase transitions of plastic crystals of spherical colloids under oscillatory shear, in particular that the continuous transition observed in spherical systems transforms into a discontinuous one.
	The latter phenomenon must be attributed to rotational-translational couplings~\cite{Lynden-Bell1994}, absent in nearly spherical systems  that apparently lead to dramatic changes in the structural and stress relaxation behavior.
	In fact, we recently showed by equilibrium BD simulations that plastic dumbbells crystals exhibit a dramatic increase of the linear shear response for high packing fractions above a critical aspect ratio of about $0.15$~\cite{Heptner2015}. 
	With respect to the sequence of shear-induced states, the type and the dynamics of the equilibrium to twinned crystal transition remains an interesting issue, which may stimulate a future study.
	The observed strong transitions have substantial implications for rheology and the yielding behavior of anisotropic colloidal crystals, as detailed by \citeauthor{Chu2015}~\cite{Chu2015}.
	The new experimental results also show a frequency and time dependency of the rheology, which we attribute to a dynamical slowing down.
	The present results also demonstrate that the thermosensitive dumbbell particles introduced before~\cite{Chu2012} serve as an excellent and versatile model system for mildly anisotropic colloids to study their equilibrium and nonequilibrium structural and phase behavior. 
	
\renewcommand{\theequation}{A\arabic{equation}}
\renewcommand{\thefigure}{A\arabic{figure}}
\setcounter{figure}{0}
\appendix{}
\section{Compilation of Rheology and Scattering}
\begin{figure}[ht!]
	\centering
	\includegraphics[]{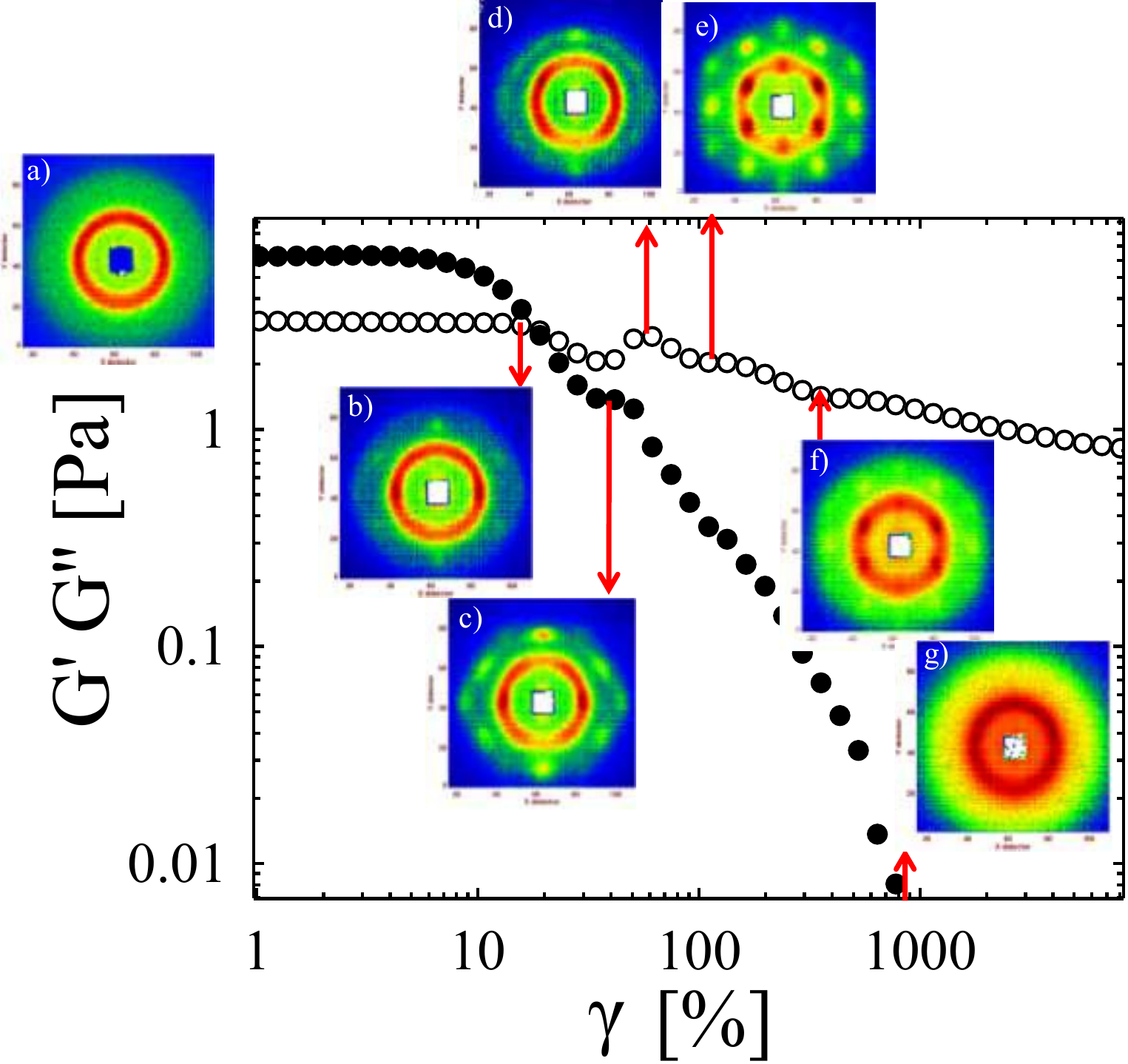}
	\caption{Yielding behavior of the hard dumbbells with $L^* \approx 0.30$ (the DPM\_b microgels) in the fully crystalline phase ($\phi_\mrm{eff}= 0.60$) in the oscillatory shear of $5 \, \mrm{Hz}$, and the corresponding 2D scattering patterns are measured by SANS. Along the dependence of $G'$ and $G''$ on various strains: (a) at rest (b) $16 \,\%$ ($\mrm{Pe}_r = 0.29$, the end of the linear regime), (c) $42.3 \,\%$ ($\mrm{Pe}_r = 0.75$ the plateau of $G'$, the minimum for $G''$), (d) $51.5 \,\%$ ($\mrm{Pe}_r = 0.90$, the maximum for $G''$), (e) $92.6 \,\%$ ($\mrm{Pe}_r = 1.62$), (f) $300 \,\%$ ($\mrm{Pe}_r = 5.37$), (g) $1000 \,\%$ ($\mrm{Pe}_r = 17.84$) }
	\label{fig:fig_s1}
\end{figure}

\bibliography{references}

\end{document}